\documentclass[letterpaper, 10pt, conference]{IEEEtran}
\IEEEoverridecommandlockouts
\usepackage{cite}
\usepackage{amsmath,amssymb,amsfonts}
\usepackage{algorithmic}
\usepackage{graphicx}
\usepackage{textcomp}
\usepackage{xcolor}
\usepackage{xcolor,colortbl}
\usepackage{threeparttable}
\usepackage{stfloats}
\usepackage{booktabs}
\usepackage{makecell}
\usepackage{stfloats} 

\definecolor{Gray}{gray}{0.85}
\def\BibTeX{{\rm B\kern-.05em{\sc i\kern-.025em b}\kern-.08em
    T\kern-.1667em\lower.7ex\hbox{E}\kern-.125emX}}
\begin{document}

\title{Enhancing Urban GNSS Positioning Reliability via Conservative Satellite Selection Using Unanimous Voting Across Multiple Machine Learning Classifiers\\
\thanks{This work was supported in part by the National Research Foundation of Korea (NRF), funded by the Korean government (Ministry of Science and ICT, MSIT), under Grant RS-2024-00358298; 
in part by the Unmanned Vehicles Core Technology Research and Development Program through the NRF and the Unmanned Vehicle Advanced Research Center (UVARC), funded by the MSIT, Republic of Korea, under Grant 2020M3C1C1A01086407;
in part by the MSIT, Korea, under the Information Technology Research Center (ITRC) support program, supervised by the Institute of Information \& Communications Technology Planning \& Evaluation (IITP), under Grant IITP-2024-RS-2024-00437494;
in part by Grant RS-2024-00407003 from the ``Development of Advanced Technology for Terrestrial Radionavigation System'' project, funded by the Ministry of Oceans and Fisheries, Republic of Korea;
and in part by the Korean government (Korea Aerospace Administration, KASA), under Grant RS-2022-NR067078.}
}

\makeatletter
\newcommand{\linebreakand}{%
  \end{@IEEEauthorhalign}
  \hfill\mbox{}\par
  \mbox{}\hfill\begin{@IEEEauthorhalign}
}
\makeatother

\author{\IEEEauthorblockN{Sanghyun Kim}
\IEEEauthorblockA{
\textit{School of Integrated Technology} \\
\textit{Yonsei University}\\
Incheon, Korea \\
sanghyun.kim@yonsei.ac.kr}
\and
\IEEEauthorblockN{Jiwon Seo${}^{*}$}
\IEEEauthorblockA{
\textit{School of Integrated Technology} \\
\textit{Yonsei University}\\
Incheon, Korea \\
jiwon.seo@yonsei.ac.kr}
{\small${}^{*}$ Corresponding author}
}

\maketitle

\begin{abstract}
In urban environments, global navigation satellite system (GNSS) positioning is often compromised by signal blockages and multipath effects caused by buildings, leading to significant positioning errors. 
To address this issue, this study proposes a robust enhancement of zonotope shadow matching (ZSM)-based positioning by employing a conservative satellite selection strategy using unanimous voting across multiple machine learning classifiers. 
Three distinct models---random forest (RF), gradient boosting decision tree (GBDT), and support vector machine (SVM)---were trained to perform line-of-sight (LOS) and non-line-of-sight (NLOS) classification based on global positioning system (GPS) signal features. 
A satellite is selected for positioning only when all classifiers unanimously agree on its classification and their associated confidence scores exceed a threshold. 
Experiments with real-world GPS data collected in dense urban areas demonstrate that the proposed method significantly improves the positioning success rate and the receiver containment rate, even with imperfect LOS/NLOS classification. 
Although a slight increase in the position bound was observed due to the reduced number of satellites used, overall positioning reliability was substantially enhanced, indicating the effectiveness of the proposed approach in urban GNSS environments.
\end{abstract}

\begin{IEEEkeywords}
global positioning system, zonotope shadow matching, machine learning, unanimous voting, urban positioning
\end{IEEEkeywords}

\section{Introduction}
Intelligent transportation systems, including autonomous vehicles, depend critically on global navigation satellite system (GNSS) to deliver reliable and accurate positioning information \cite{Lee22:Urban, Ma20:Articial, Jia21:Ground, Chen11:Real, Lee22:Optimal}. 
However, in urban environments, satellite signals can be blocked or reflected by obstacles such as buildings, resulting in the reception of non-line-of-sight (NLOS) signals \cite{Kim22:Machine, Kim23:Single, Lee23:Seamless, Zhu18, Park23:Detection, Park24:CSAC, Lee25:Reducing}. 
These cause significant errors in range measurements and leads to inaccurate position estimation \cite{MacGougan02}. 

Among the various techniques \cite{Moon24:HELPS, Lee23:Performance_Comparison, Kim23:Low, Lee23:Performance_Evaluation, Lee22:Evaluation} proposed to improve positioning performance in urban environments, 3D mapping-aided (3DMA) methods that utilize 3D city models have emerged as a promising solution \cite{Kumar14, Zhong22:Multi, Lee22:Nonlinear, Lee23:Nonlinear}. 
A representative technique is shadow matching \cite{Groves11:Shadow}, which leverages GNSS shadows—areas where line-of-sight (LOS) signals cannot reach. 
Shadow matching estimates the receiver's position by finding the location where the predicted LOS/NLOS results based on GNSS signal features best match those predicted using the 3D city model. 
Building on its initial success, various advanced methodologies have been proposed to further improve shadow matching under diverse urban conditions \cite{Wang12, Adjrad18:Intelligent, Wang13:GNSS}. 
Despite these advancements, most existing implementations still adopt a grid-based manner, which may impose limitations in terms of adaptability and computational efficiency. 
In contrast, shadow matching can alternatively be formulated in a set-based framework, where the user’s location is represented as being either within a shadow region—modeled as a 2D set or polygon—or outside it, in the complement space. 
This set-based perspective eliminates the need for discretizing the search space into a predefined grid of position candidates, which is a common requirement in conventional grid-based approaches. 
It also facilitates straightforward verification of whether the estimated position lies within user-defined safety areas \cite{Kim25:Set}. 

One notable example of a set-based implementation is the zonotope shadow matching (ZSM) method \cite{Bhamidipati22:Set, Kim24:Performance, Kim25:Set}, which extends the shadow matching framework by modeling all geometric components as mathematical sets, including 3D representations of buildings. 
Leveraging the mathematical structure of constrained zonotopes, ZSM allows for efficient real-time computation of shadow regions through fast vector-based operations. 
The method iteratively refines the area of interest (AOI) by performing set intersection and subtraction, enabling accurate localization without relying on a predefined grid. 

While ZSM effectively addresses key challenges in shadow matching, such as resolution, computational efficiency, and certifiable error bounds, it still faces certain limitations \cite{Neamati22:Set}. 
One of the main limitations is its reliance on accurate LOS/NLOS classification. 
ZSM refines the AOI by treating LOS satellites as indicators that the receiver is outside the shadow region, and NLOS satellites as indicators that it is inside. 
The classifications of satellite signals as LOS or NLOS are typically provided by artificial intelligence (AI)-based models trained on GNSS signal features \cite{Adjrad17:Enhancing, Kim23:Machine, Zhu25, Jeong24:Quantum}. 
However, when misclassification occurs, the AOI may be refined toward areas that do not contain the true receiver location, potentially leading to localization failure, as illustrated in Fig.~\ref{fig:Misclassification}. 
Such failures pose significant risks in urban environments, where reliable positioning is often essential for safety and robustness. 
Although AI-based LOS/NLOS classifiers generally achieve around 80\% accuracy \cite{Xu24}, their performance can be improved with additional sensors such as fish-eye cameras or LiDARs---yet these sensors are not universally available to all users. 

\begin{figure*}
  \centering
  \includegraphics[width=0.8\linewidth]{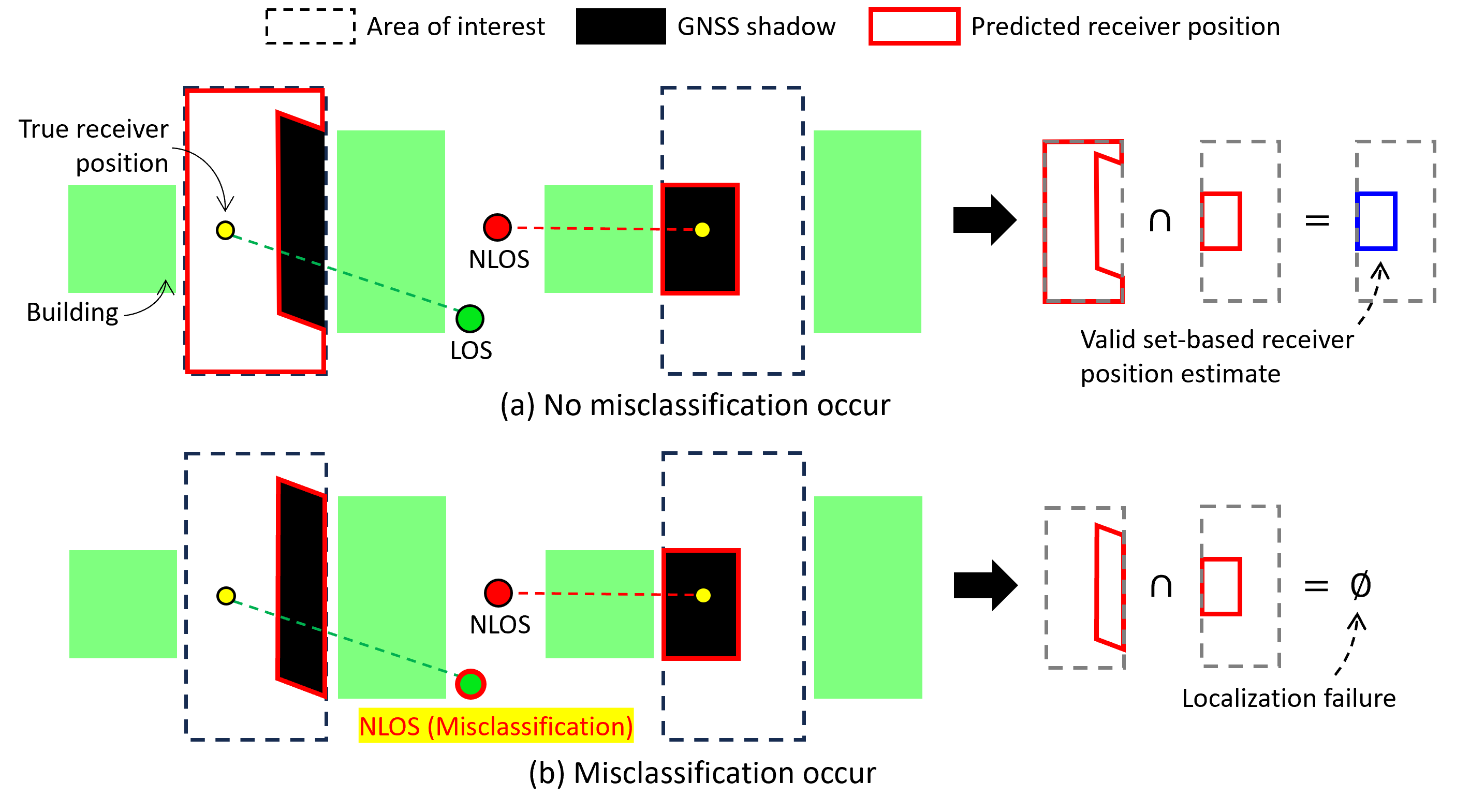}
  \caption{Positioning examples of the ZSM algorithm. (a) A case without misclassification, where all signals are correctly classified, resulting in a valid receiver position estimate. (b) A case with misclassification, where a LOS signal is incorrectly classified as NLOS. As a result, the predicted receiver positions from individual satellites do not intersect, leading to localization failure.}
  \label{fig:Misclassification}
\end{figure*}

In this study, we propose a method that enhances the robustness of the ZSM algorithm by improving the likelihood of producing a valid receiver position estimate, even when relying on imperfect AI-based LOS/NLOS classification models. 
The core idea is to minimize the influence of misclassified satellites by reducing the number of erroneous inputs used in the positioning process. 
To this end, we introduce a conservative selection strategy in which a satellite is used in the position estimation process only when all three independently trained machine learning classifiers agree on its LOS/NLOS label, satisfying a unanimous voting condition. 
To further enhance the reliability of the selected signals, we additionally require that each classifier assigns a confidence probability exceeding a predefined threshold. 
This ensures that only highly reliable satellites contribute to the positioning solution. 
We evaluate the proposed method using real-world global positioning system (GPS) signal data collected in urban environments. 
Experimental results demonstrate that our approach significantly increases the positioning success rate—defined as the proportion of epochs in which ZSM yields a valid position estimate—compared to using individual classifiers. 

\section{Method}

\begin{figure*}
  \centering
  \includegraphics[width=0.8\linewidth]{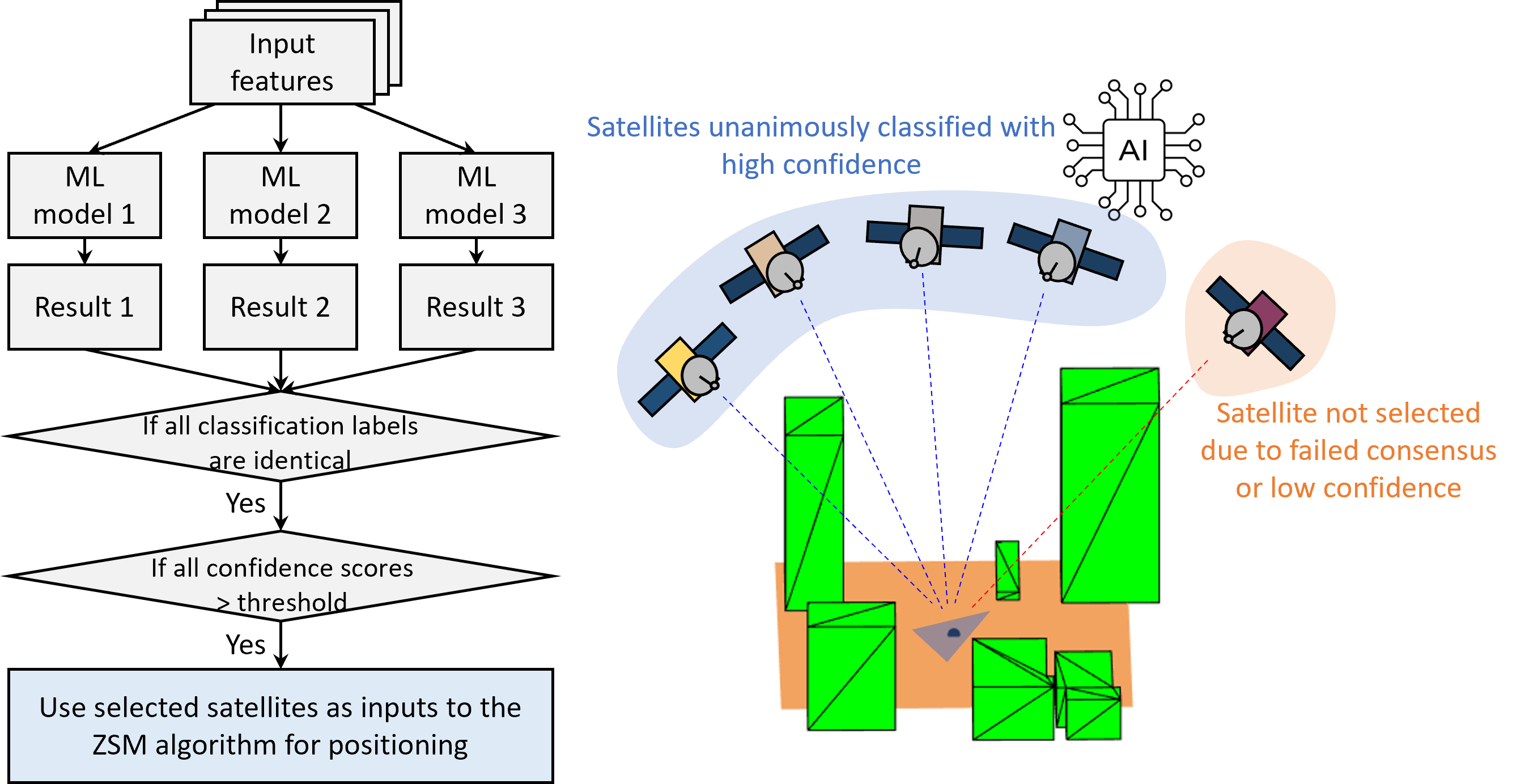}
  \caption{Overview of the proposed conservative satellite selection algorithm for robust ZSM-based positioning.}
  \label{fig:Proposed}
\end{figure*}

\subsection{Algorithm Overview}

An overview of the proposed algorithm---a conservative satellite selection method aimed at improving the robustness of ZSM---is illustrated in Fig.~\ref{fig:Proposed}. 
For each visible satellite observed within a single epoch, LOS/NLOS classification is performed using three machine learning models, each based on a distinct algorithm. 
A satellite is included in the position estimation only if all classifiers unanimously agree on its LOS/NLOS label and all of their confidence scores exceed a predefined threshold. 
The confidence score from each model is defined as the highest class probability it assigns to the sample, that is, the probability associated with either LOS or NLOS classification based on the corresponding satellite's measurement features.
In our implementation, the threshold was empirically determined to be 0.7 based on experimental results. 
Satellites that do not meet both criteria are considered potentially unreliable and are excluded from the positioning process.

\subsection{Feature Selection and Machine Learning Algorithms}

In this study, three features were used for LOS/NLOS classification: satellite elevation angle, carrier-to-noise-density ratio ($C/N_0$), and pseudorange residual. 
This combination, which includes features commonly regarded as essential for distinguishing LOS from NLOS signals, has been employed in various previous studies \cite{Xu24, Sun20}. 
The characteristics of each feature are as follows:

\begin{itemize}
\item Satellite elevation angle: Signals with higher elevation angles are more likely to be LOS, whereas those with lower elevation angles are more susceptible to obstruction by buildings and, therefore, more likely to be NLOS.

\item $C/N_0$: NLOS signals typically exhibit lower $C/N_0$ values due to additional propagation losses from reflections off buildings, as compared to LOS signals.

\item Pseudorange residual: The pseudorange residual represents the difference between the measured pseudorange and the geometric distances computed from the estimated receiver position to each satellite. It is derived as follows: 
\begin{equation}
\boldsymbol{\delta} = \tilde{\boldsymbol{\rho}} - \mathbf{G} \hat{\mathbf{p}}
\end{equation}
\begin{equation}
\hat{\mathbf{p}} = \left( \mathbf{G}^\top \mathbf{G} \right)^{-1} \mathbf{G}^\top \tilde{\boldsymbol{\rho}}
\end{equation}
where $\boldsymbol{\delta}$ is the pseudorange residual vector, $\tilde{\boldsymbol{\rho}}$ is the pseudorange measurements, $\hat{\mathbf{p}}$ is the estimated receiver state (including the three-dimensional position and receiver clock bias), and $\mathbf{G}$ is the geometry matrix composed of unit LOS vectors between the receiver and satellites. Generally, larger pseudorange residuals tend to be associated with a higher likelihood of NLOS signal reception.
\end{itemize}

The three machine learning algorithms used in this study are random forest (RF) \cite{Breiman01}, gradient boosting decision tree (GBDT) \cite{Friedman01}, and support vector machine (SVM) \cite{Cortes95}. The characteristics of each algorithm are as follows:

\begin{itemize}
\item Random forest (RF): RF is an ensemble learning method that constructs multiple decision trees during training and outputs the majority vote for classification tasks. It is known for its robustness to noise and overfitting, especially when dealing with high-dimensional or imbalanced datasets.

\item Gradient boosting decision tree (GBDT): GBDT is a sequential ensemble technique that builds decision trees in a stage-wise manner, with each new tree correcting the errors of its predecessors. It achieves high accuracy by minimizing a loss function through gradient descent.

\item Support vector machine (SVM): SVM is a supervised learning algorithm that seeks to find the optimal hyperplane that separates data points of different classes with the maximum margin. It is effective in high-dimensional spaces and can model non-linear relationships using kernel functions. 
\end{itemize}

\section{GPS Signal Collection and Labeling}

To evaluate the performance of the proposed algorithm, GPS signal data were collected in a dense urban environment, as illustrated in Fig.~\ref{fig:Signal}. 
A vehicle equipped with an Antcom 3G1215RL-AA-XT-1 dual-polarized antenna and a NovAtel PwrPak7 receiver was driven through the area, and GPS signals were recorded at 1~Hz. 
To obtain a high-precision ground-truth trajectory, a NovAtel GNSS/INS system---consisting of a GPS-703-GGG antenna, SPAN-SE, and UIMU-H58---was also mounted on the vehicle. 
The GNSS/INS data were sampled at 1~Hz and post-processed using NovAtel Inertial Explorer software to generate accurate reference trajectories.

A ray-tracing algorithm based on a 3D city model was employed to label the collected GPS signals as LOS or NLOS. 
A signal was labeled as NLOS if the vector connecting the satellite and the true receiver position intersected a building; otherwise, it was labeled as LOS. 
The 3D city model used for this labeling process was a commercial product provided by ONEGEO \cite{ONEGEO}. 

To evaluate the positioning success rate of the ZSM algorithm, a specific section of the road was designated as the target road. 
Data collected from all other sections were used to construct the training dataset, which was used to train various machine learning models. 
The trained models' classification accuracy was evaluated using a test dataset consisting solely of samples collected from the target road. 
The composition of LOS and NLOS data samples in the training and test datasets is summarized in Table~\ref{table:LabelResult}.

\begin{figure}
  \centering
  \includegraphics[width=0.75\linewidth]{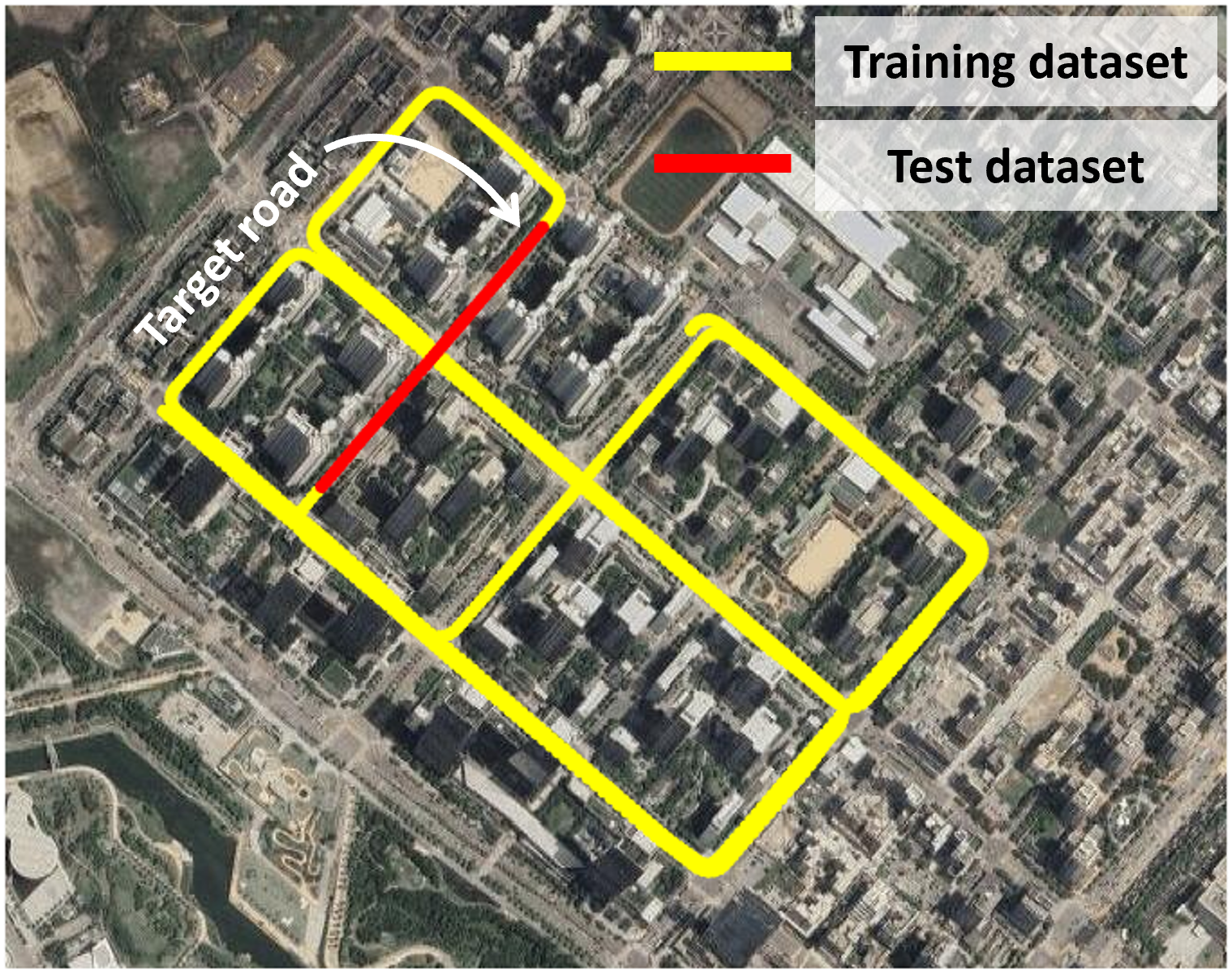}
  \caption{Urban area in Songdo, Incheon, South Korea, where GPS signals were collected.}
  \label{fig:Signal}
\end{figure}

\begin{table}
\centering
\caption{Number of data samples in the training and test datasets}
\renewcommand{\arraystretch}{1.5}
\begin{tabular}{l||cc}\hline
  & Training dataset & Test dataset \\
 \hline
 LOS samples & 10,800 & 705 \\
 NLOS samples & 1,677 & 212 \\
 \hline
 Total samples & 12,477 & 917 \\
 \hline
\end{tabular}
\label{table:LabelResult}
\end{table}

\section{Results and Discussions}

\subsection{Signal Classification Results}

Table~\ref{table:ClassificationResult} summarizes the classification accuracy of the three trained machine learning models evaluated on the test dataset. 
The classification accuracies achieved were 80.9\% for the RF model, 85.9\% for the GBDT model, and 83.2\% for the SVM model.  
Fig.~\ref{fig:Visible} shows the number of visible GPS satellites per epoch along the target road, where a total of 146 epochs were collected, with an average of 6.28 satellites per epoch. 
The average number of misclassified satellites per epoch was 1.20, 0.88, and 1.05 for Models 1 (RF), 2 (GBDT), and 3 (SVM), respectively. 

\begin{table}
\centering 
\centering
\caption{Classification accuracy of the three trained machine learning models on the test dataset}
\renewcommand{\arraystretch}{1.5}
\begin{tabular}{|>{\centering\arraybackslash}m{1.8cm}|>{\centering\arraybackslash}m{1.5cm}|
>{\centering\arraybackslash}m{1.5cm}|
>{\centering\arraybackslash}m{1.5cm}|}\hline
 & Model 1 (RF) & Model 2 (GBDT) & Model 3 (SVM) \\
 \hline
 Classification accuracy & 80.9\% & 85.9\% & 83.2\% \\
 \hline
\end{tabular}
\label{table:ClassificationResult}
\end{table}

\begin{figure}
  \centering
  \includegraphics[width=0.70\linewidth]{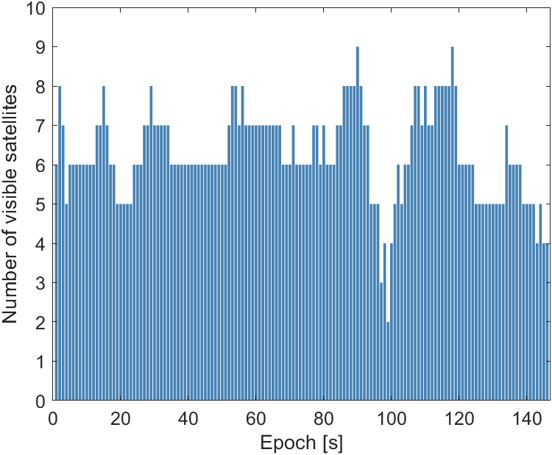}
  \caption{Number of visible GPS satellites per epoch along the target road.}
  \label{fig:Visible}
\end{figure}

\begin{table}
\centering
\caption{Comparison of positioning success rates}
\renewcommand{\arraystretch}{1.5}
\begin{tabular}{>{\centering\arraybackslash}m{5cm} >{\centering\arraybackslash}m{3cm}}
\toprule
\textbf{Method} & \textbf{Positioning success rate} \\
\midrule
Model 1 (RF) & 97.3\% (142/146) \\
\addlinespace
Model 2 (GBDT) & 91.8\% (134/146) \\
\addlinespace
Model 3 (SVM) & 95.9\% (140/146) \\
\addlinespace
\makecell{Unanimous voting \\ (Proposed)} & 99.3\% (145/146) \\
\addlinespace
\makecell{Unanimous voting w/ confidence threshold \\ (Proposed)} & 100\% (146/146) \\
\bottomrule
\end{tabular}
\label{table:PositioningSuccessRateResult}
\end{table}

By applying the unanimous voting strategy, 86.3\% of all observed satellite signals were selected unanimously by all three models. 
When the confidence score threshold condition was additionally applied, 70.9\% of the satellite signals satisfied both the unanimous voting and confidence criteria, while the remaining 29.1\% were excluded from the positioning process. 
Among the selected satellites, 90.8\% were correctly classified, resulting in an average of 0.41 misclassified satellites per epoch, which corresponds to a reduction of approximately 53-65\% compared to using individual models.

\subsection{Positioning Success Rate}

The ZSM-based positioning performance was evaluated across the 146 epochs using both individual models and the proposed conservative selection strategy. 
As shown in Table~\ref{table:PositioningSuccessRateResult},  the positioning success rates, defined as the proportion of epochs in which ZSM provides a valid position estimate, achieved by Models 1, 2, and 3 were 97.3\%, 91.8\%, and 95.9\%, respectively. 
In contrast, the unanimous voting approach improved the success rate to 99.3\%, and further applying the confidence threshold led to a 100\% success rate.  
These results clearly demonstrate that conservatively selecting satellites based on consensus among multiple classifiers significantly enhances the robustness and reliability of ZSM positioning. 

\subsection{Analysis of Positioning Results}

\begin{table*}
\centering
\caption{Comparison of receiver containment rates and position bounds}
\renewcommand{\arraystretch}{1.5}
\begin{tabular}{>{\centering\arraybackslash}m{6.0cm} >{\centering\arraybackslash}m{4.0cm} >{\centering\arraybackslash}m{3.0cm} >{\centering\arraybackslash}m{3.0cm}} 
\toprule
\textbf{Method} & \textbf{Receiver containment rate} & \makecell{\textbf{Cross-street}\\\textbf{position bound}} & \makecell{\textbf{Along-street}\\\textbf{position bound}} \\
\midrule
Model 1 (RF) & 26.0\% (38/146) & 38.6 m & 52.9 m \\
\addlinespace
Model 2 (GBDT) & 41.1\% (60/146) & 40.2 m & 51.9 m \\
\addlinespace
Model 3 (SVM) & 42.5\% (62/146) & 41.6 m & 59.0 m \\
\addlinespace
\makecell{Unanimous voting\\(Proposed)} & 50.7\% (74/146) & 45.8 m & 79.1 m \\
\addlinespace
\makecell{Unanimous voting w/ confidence threshold\\(Proposed)} & 61.0\% (89/146) & 52.3 m & 108.7 m \\
\bottomrule
\end{tabular}
\label{table:PositioningResult2}
\end{table*}

We further analyzed the positioning results obtained using the ZSM algorithm for each method. 
In particular, we examined the receiver containment rate, which is defined as the proportion of epochs during which the estimated receiver position set successfully encompasses the true receiver position. 
Ensuring that the true receiver position falls within the estimated region is critical for maintaining the reliability of the positioning result. 
In addition, we evaluated the size of the estimated receiver position region, referred to as the position bound \cite{Bhamidipati22:Set}. 
Reducing the position bound is desirable for enhancing positioning precision and achieving a tighter localization of the receiver. 
The position bounds were separately measured in the cross-street and along-street directions. 

Table~\ref{table:PositioningResult2} compares the receiver containment rate and position bound for each method.
Models 1, 2, and 3 achieved receiver containment rates of 26.0\%, 41.1\%, and 42.5\%, respectively.
Since a misclassification of even a single satellite at a given epoch can cause the estimated receiver position to fail to encompass the true receiver position, the lowest receiver containment rate was observed when using Model 1, which had the lowest classification accuracy.
By applying the unanimous voting method, where only unanimously selected satellites are used for positioning instead of using all satellites, the number of epochs without misclassification increased, resulting in a higher probability (50.7\%) that the true receiver position would be included in the positioning result.
When a confidence threshold is additionally considered to select satellites more conservatively, the receiver containment rate further improves to 61.0\%.
Therefore, by applying the proposed technique, it was confirmed that a more reliable set-based estimated receiver position could be obtained. 

However, in terms of the position bound, since the number of satellites used for positioning decreases, the estimated receiver position set was not sufficiently refined.
As shown in Table~\ref{table:PositioningResult2}, both the cross-street and along-street position bounds became larger when satellites were conservatively selected, compared to using all available satellites as in Models 1, 2, and 3. 
This phenomenon is likely more pronounced in this study, as only GPS satellites were used, resulting in a limited number of visible satellites. 
It is expected that if multi-constellation satellites were employed to increase the total number of satellites, the issue of enlarged position bounds could be mitigated. 

Fig.~\ref{fig:Position} further illustrates the positioning results at the epoch 56. 
Fig.~\ref{fig:Position}(a) shows the result obtained using the proposed method with unanimous voting and a confidence threshold. 
Out of eight visible satellites, five conservatively selected satellites were used, and no misclassified satellites were included, resulting in a reliable position set that encompassed the true receiver position. 
In contrast, Fig.~\ref{fig:Position}(b) presents the result using Model 1, where all eight satellites were used, including one misclassified satellite, leading to a position set that failed to contain the true position. 
Nonetheless, the proposed conservative satellite selection method showed limitations in reducing the position bound. 
As previously mentioned, this issue could be alleviated by employing a multi-constellation system to increase the number of available satellites. 

\begin{figure}
  \centering
  \includegraphics[width=1.0\linewidth]{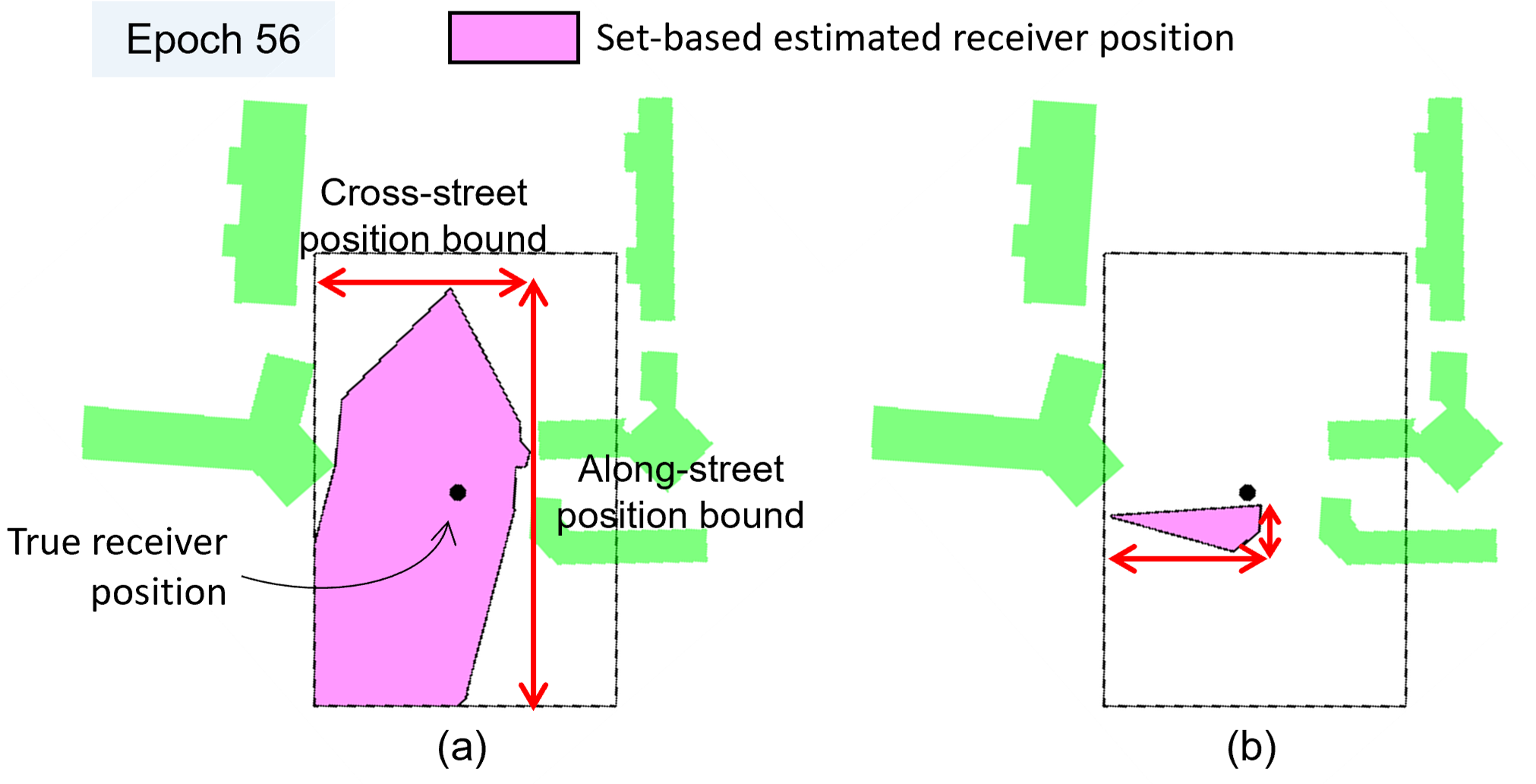}
  \caption{Comparison of positioning results for epoch 56. (a) shows the result of the proposed method based on unanimous voting with a confidence threshold, where five satellites were selected and no misclassified satellites were included. (b) presents the result of model 1, which used eight satellites including one misclassified satellite.}
  \label{fig:Position}
\end{figure}

\section{Conclusion}
In this study, a conservative satellite selection method was proposed to improve the robustness of set-based GNSS positioning in urban environments, with a particular focus on the ZSM framework. 
By applying unanimous voting across three independently trained machine learning classifiers and by imposing a confidence threshold, the proposed method selectively utilizes only highly reliable satellites for positioning. 
Experimental validation using real-world GPS signals showed that the proposed strategy effectively reduced the number of misclassified signals, achieving a 100\% positioning success rate and significantly improving the receiver containment rate compared to conventional single-model approaches. 
Although the reduced number of satellites led to an expansion of the position bound, this limitation is expected to be alleviated through the integration of multi-constellation GNSS signals, which is planned for future work. 
Overall, the proposed conservative selection framework demonstrates a practical and effective way to achieve safer and more reliable GNSS positioning in challenging urban environments. 

\bibliographystyle{IEEEtran}
\bibliography{mybibfile, IUS_publications}

\begin{thebibliography}{10}
\providecommand{\url}[1]{#1}
\csname url@samestyle\endcsname
\providecommand{\newblock}{\relax}
\providecommand{\bibinfo}[2]{#2}
\providecommand{\BIBentrySTDinterwordspacing}{\spaceskip=0pt\relax}
\providecommand{\BIBentryALTinterwordstretchfactor}{4}
\providecommand{\BIBentryALTinterwordspacing}{\spaceskip=\fontdimen2\font plus
\BIBentryALTinterwordstretchfactor\fontdimen3\font minus \fontdimen4\font\relax}
\providecommand{\BIBforeignlanguage}[2]{{%
\expandafter\ifx\csname l@#1\endcsname\relax
\typeout{** WARNING: IEEEtran.bst: No hyphenation pattern has been}%
\typeout{** loaded for the language `#1'. Using the pattern for}%
\typeout{** the default language instead.}%
\else
\language=\csname l@#1\endcsname
\fi
#2}}
\providecommand{\BIBdecl}{\relax}
\BIBdecl

\bibitem{Lee22:Urban}
H.~Lee, J.~Seo, and Z.~Kassas, ``Urban road safety prediction: A satellite navigation perspective,'' \emph{IEEE Intell. Transp. Syst. Mag.}, vol.~14, no.~6, pp. 94--106, Nov.-Dec. 2022.

\bibitem{Ma20:Articial}
Y.~Ma, Z.~Wang, H.~Yang, and L.~Yang, ``Artificial intelligence applications in the development of autonomous vehicles: a survey,'' \emph{IEEE/CAA J. Automatica Sinica}, vol.~7, no.~2, pp. 315--329, 2020.

\bibitem{Jia21:Ground}
M.~Jia, H.~Lee, J.~Khalife, Z.~M. Kassas, and J.~Seo, ``Ground vehicle navigation integrity monitoring for multi-constellation {GNSS} fused with cellular signals of opportunity,'' in \emph{Proc. IEEE ITSC}, 2021, pp. 3978--3983.

\bibitem{Chen11:Real}
Y.-H. Chen, J.-C. Juang, D.~{De Lorenzo}, J.~Seo, S.~Lo, P.~Enge, and D.~Akos, ``Real-time dual-frequency ({L1/L5}) {GPS/WAAS} software receiver,'' in \emph{Proc. ION GNSS}, 2011, pp. 767--774.

\bibitem{Lee22:Optimal}
H.~Lee, S.~Pullen, J.~Lee, B.~Park, M.~Yoon, and J.~Seo, ``Optimal parameter inflation to enhance the availability of single-frequency {GBAS} for intelligent air transportation,'' \emph{IEEE Trans. Intell. Transp. Syst.}, vol.~23, no.~10, pp. 17\,801--17\,808, Oct. 2022.

\bibitem{Kim22:Machine}
S.~Kim, J.~Byun, and K.~Park, ``Machine learning-based {GPS} multipath detection method using dual antennas,'' in \emph{Proc. ASCC}, May 2022, pp. 691--695.

\bibitem{Kim23:Single}
S.~Kim, S.~Park, and J.~Seo, ``Single antenna based {GPS} signal reception condition classification using machine learning approaches,'' \emph{J. Position. Navig. Timing}, vol.~12, no.~2, pp. 149--155, 2023.

\bibitem{Lee23:Seamless}
Y.~Lee, Y.~Hwang, J.~Y. Ahn, J.~Seo, and B.~Park, ``Seamless accurate positioning in deep urban area based on mode switching between {DGNSS} and multipath mitigation positioning,'' \emph{IEEE Trans. Intell. Transp. Syst.}, vol.~24, no.~6, pp. 5856--5870, Jun. 2023.

\bibitem{Zhu18}
N.~Zhu, J.~Marais, D.~Bétaille, and M.~Berbineau, ``{GNSS} position integrity in urban environments: A review of literature,'' \emph{IEEE Trans. Intell. Transp. Syst.}, vol.~19, no.~9, pp. 2762--2778, 2018.

\bibitem{Park23:Detection}
S.~Park, T.~Kang, S.~Lee, and J.~H. Rhee, ``Detection of pedestrian turning motions to enhance indoor map matching performance,'' in \emph{Proc. ICTC}, 2023, pp. 393--398.

\bibitem{Park24:CSAC}
S.~Park and J.~H. Rhee, ``{CSAC} drift modeling considering {GPS} signal quality in the case of {GPS} signal unavailability,'' in \emph{Proc. ICCAS}, 2024, pp. 187--192.

\bibitem{Lee25:Reducing}
H.~Lee and J.~Seo, ``Reducing computational complexity of rigidity-based {UAV} trajectory optimization for real-time cooperative target localization,'' in \emph{Proc. ION ITM}, Jan 2025, pp. 80--87.

\bibitem{MacGougan02}
G.~MacGougan, G.~Lachapelle, R.~Klukas, K.~Siu, L.~Garin, J.~Shewfelt, and G.~Cox, ``Performance analysis of a stand-alone high-sensitivity receiver,'' \emph{GPS Solut.}, vol.~6, no.~3, pp. 179--195, 2002.

\bibitem{Moon24:HELPS}
H.~Moon, H.~Park, and J.~Seo, ``{HELPS} for emergency location service: Hyper-enhanced local positioning system,'' \emph{IEEE Wirel. Commun.}, vol.~31, no.~4, pp. 276--282, 2024.

\bibitem{Lee23:Performance_Comparison}
H.~Lee and J.~Seo, ``Performance comparison of numerical optimization algorithms for {RSS}-{TOA}-based target localization,'' in \emph{Proc. IEEE VTC}, Jun. 2023, pp. 1--6.

\bibitem{Kim23:Low}
W.~Kim and J.~Seo, ``Low-cost {GNSS} simulators with wireless clock synchronization for indoor positioning,'' \emph{IEEE Access}, vol.~11, pp. 55\,861--55\,874, 2023.

\bibitem{Lee23:Performance_Evaluation}
H.~Lee and J.~Seo, ``Performance evaluation and hybrid application of the greedy and predictive {UAV} trajectory optimization methods for localizing a target mobile device,'' in \emph{Proc. ION ITM}, Jan. 2023, pp. 161--171.

\bibitem{Lee22:Evaluation}
H.~Lee, T.~Kang, S.~Jeong, and J.~Seo, ``Evaluation of {RF} fingerprinting-aided {RSS}-based target localization for emergency response,'' in \emph{Proc. IEEE VTC}, Jun. 2022, pp. 1--7.

\bibitem{Kumar14}
R.~Kumar and M.~G. Petovello, ``A novel {GNSS} positioning technique for improved accuracy in urban canyon scenarios using {3D} city model,'' in \emph{Proc. ION GNSS+}, Sep. 2014, pp. 2139--2148.

\bibitem{Zhong22:Multi}
Q.~Zhong and P.~D. Groves, ``Multi-epoch {3D}-mapping-aided positioning using {Bayesian} filtering techniques,'' \emph{Navig. J. Inst. Navig.}, vol.~69, no.~2, 2022.

\bibitem{Lee22:Nonlinear}
Y.~Lee and B.~Park, ``Nonlinear regression-based {GNSS} multipath modelling in deep urban area,'' \emph{Mathematics}, vol.~10, no.~3, pp. 1--15, 2022.

\bibitem{Lee23:Nonlinear}
Y.~Lee, P.~Wang, and B.~Park, ``Nonlinear regression-based {GNSS} multipath dynamic map construction and its application in deep urban areas,'' \emph{IEEE Trans. Intell. Transp. Syst.}, vol.~24, no.~5, pp. 5082--5093, 2023.

\bibitem{Groves11:Shadow}
P.~D. Groves, ``Shadow matching: A new {GNSS} positioning technique for urban canyons,'' \emph{J. Navig.}, vol.~64, no.~3, pp. 417--430, 2011.

\bibitem{Wang12}
L.~Wang, P.~D. Groves, and M.~K. Ziebart, ``Multi-constellation {GNSS} performance evaluation for urban canyons using large virtual reality city models,'' \emph{J. Navig.}, vol.~65, pp. 459--476, 2012.

\bibitem{Adjrad18:Intelligent}
M.~Adjrad and P.~D. Groves, ``Intelligent urban positioning: Integration of shadow matching with {3D}-mapping-aided {GNSS} ranging,'' \emph{J. Navig.}, vol.~71, no.~1, pp. 1--20, 2018.

\bibitem{Wang13:GNSS}
L.~Wang, P.~D. Groves, and M.~K. Ziebart, ``{GNSS} shadow matching: Improving urban positioning accuracy using a {3D} city model with optimized visibility prediction scoring,'' \emph{Navig. J. Inst. Navig.}, vol.~60, no.~3, pp. 195--207, 2013.

\bibitem{Kim25:Set}
S.~Kim and J.~Seo, ``Set-based position ambiguity reduction method for zonotope shadow matching in urban areas using estimated multipath errors,'' in \emph{Proc. ION ITM}, Jan 2025, pp. 1--10.

\bibitem{Bhamidipati22:Set}
S.~Bhamidipati, S.~Kousik, and G.~Gao, ``Set-valued shadow matching using zonotopes for {3D}-map-aided {GNSS} localization,'' \emph{Navig. J. Inst. Navig.}, vol.~69, no.~4, 2022.

\bibitem{Kim24:Performance}
S.~Kim and J.~Seo, ``Performance analysis of zonotope shadow matching algorithm according to various urban environments,'' \emph{J. Position. Navig. Timing}, vol.~13, no.~3, pp. 215--220, 2024.

\bibitem{Neamati22:Set}
D.~Neamati, S.~Bhamidipati, and G.~Gao, ``Set-based ambiguity reduction in shadow matching with iterative {GNSS} pseudoranges,'' in \emph{Proc. ION GNSS+}, Sep. 2022, pp. 1093--1107.

\bibitem{Adjrad17:Enhancing}
M.~Adjrad and P.~D. Groves, ``Enhancing least squares {GNSS} positioning with {3D} mapping without accurate prior knowledge,'' \emph{Navig. J. Inst. Navig.}, vol.~64, no.~1, pp. 75--91, 2017.

\bibitem{Kim23:Machine}
S.~Kim and J.~Seo, ``Machine-learning-based classification of {GPS} signal reception conditions using a dual-polarized antenna in urban areas,'' in \emph{Proc. IEEE/ION PLANS}, Apr. 2023, pp. 113--118.

\bibitem{Zhu25}
N.~Zhu, R.~He, and Z.~Wang, ``{CarNet}: A generative convolutional neural network-based line-of-sight/non-line-of-sight classifier for global navigation satellite systems by transforming multivariate time-series data into images,'' \emph{Eng. Appl. Artif. Intell.}, vol. 145, no. 110160, pp. 1--16, 2025.

\bibitem{Jeong24:Quantum}
S.~Jeong, S.~Kim, and J.~Seo, ``Quantum support vector machine-based classification of {GPS} signal reception conditions,'' in \emph{Proc. QCE}, 2024, pp. 530--531.

\bibitem{Xu24}
P.~Xu, G.~Zhang, B.~Yang, and L.-T. Hsu, ``Machine learning in {GNSS} multipath/{NLOS} mitigation: Review and benchmark,'' \emph{IEEE Aerosp. Electron. Syst. Mag.}, vol.~39, no.~9, pp. 26--44, 2024.

\bibitem{Sun20}
R.~Sun, G.~Wanga, W.~Zhang, L.-T. Hsu, and W.~Y. Ochieng, ``A gradient boosting decision tree based {GPS} signal reception classification algorithm,'' \emph{Appl. Soft Comput. J.}, vol.~86, pp. 1--12, 2020.

\bibitem{Breiman01}
L.~Breiman, ``Random forests,'' \emph{Mach. Learn.}, vol.~45, pp. 5--32, 2001.

\bibitem{Friedman01}
J.~H. Friedman, ``Greedy function approximation: A gradient boosting machine,'' \emph{Ann. Stat.}, vol.~29, no.~5, p. 1189–1232, 2001.

\bibitem{Cortes95}
C.~Cortes and V.~Vapnik, ``Support-vector networks,'' \emph{Mach. Learn.}, vol.~20, no.~3, pp. 273--297, 1995.

\bibitem{ONEGEO}
ONEGEO. \url{https://onegeo.co/}.

\end{thebibliography}

\end{document}